\documentclass[runningheads]{llncs}
\usepackage{graphicx}
\usepackage{mathtools}
\usepackage{booktabs}
\usepackage{dirtytalk}
\begin{document}
%
%\title{Contribution Title\thanks{Supported by organization x.}}
\title{Challenging Current Semi-Supervised Anomaly Segmentation Methods for Brain MRI}
\titlerunning{Challenging Current Semi-Supervised Anomaly Segmentation Methods}
% If the paper title is too long for the running head, you can set
% an abbreviated paper title here
%
%\author{First Author\inst{1}\orcidID{0000-1111-2222-3333} \and
%Second Author\inst{2,3}\orcidID{1111-2222-3333-4444} \and
%Third Author\inst{3}\orcidID{2222--3333-4444-5555}}
\author{Felix Meissen\inst{1,2} \and
Georgios Kaissis\inst{1,2,3} \and
Daniel Rueckert\inst{1,2,3}}
\authorrunning{F. Meissen et al.}
% First names are abbreviated in the running head.
% If there are more than two authors, 'et al.' is used.
%
\institute{Technical University of Munich (TUM), Munich, Germany \and
Klinikum Rechts der Isar, Munich, Germany \and
Imperial College London, UK
\email{\{felix.meissen,g.kaissis,daniel.rueckert\}@tum.de}}
%}
%
\maketitle              % typeset the header of the contribution
\begin{abstract}
In this work, we tackle the problem of Semi-Supervised Anomaly Segmentation (SAS) in Magnetic Resonance Images (MRI) of the brain, which is the task of automatically identifying pathologies in brain images.
Our work challenges the effectiveness of current Machine Learning (ML) approaches in this application domain by showing that thresholding Fluid-attenuated inversion recovery (FLAIR) MR scans provides better anomaly segmentation maps than several different ML-based anomaly detection models.
Specifically, our method achieves better Dice similarity coefficients and Precision-Recall curves than the competitors on various popular evaluation data sets for the segmentation of tumors and multiple sclerosis lesions.
\footnote{Code available under: \url{https://github.com/FeliMe/brain\_sas\_baseline}}

\end{abstract}
\section{Introduction}
The medical imaging domain is characterized by large amounts of data, but their usability for machine learning is limited due to the challenges in sharing the data and the difficulties in obtaining labels, which requires annotations by expert radiologists and is time-consuming and costly. Especially pixel- or voxel-wise segmentation of different diseases in medical images is a tedious task.
semi-supervised machine learning seems like a natural fit to gain insights into the analysis of medical images for diagnosis as it requires no annotations and can easily utilize the large amounts of data available.
Especially valuable in this domain is Semi-Supervised Anomaly Segmentation (SAS). Here, unlabelled imaging data is used to build a system that can automatically detect anything that deviates from the \say{norm} when presented with unseen data. In medical images, this technique is particularly helpful as anomalies here often indicate morphological manifestations of pathology.

Recently, SAS achieved impressive successes in automatic industrial defect detection \cite{SSIMAE,mvtecVAE,mvtecVAE2,patchSVDD} on the MVTec-AD data set \cite{mvtecad}. In the medical imaging domain, most works have focused on the detection of pathologies in brain images. Here, mostly autoencoder-based approaches have been applied so far \cite{BrainLesionAE,BrainVAE,AnoVAEGAN,contextVAE,GMVAE,constrainedAE,ComparativeStudy}. These techniques use only images from healthy subjects as training data to learn the distribution of \say{normal} brain anatomies. During inference, most of the approaches compute a so-called anomaly map as the pixel-wise residual between the input image and a predicted \say{normal} version of the same image generated by the model, that is closer to the training distribution.
Common anomaly types in brain MRI are tumors and lesions from specific diseases such as multiple sclerosis (MS). In fact, all of the aforementioned works evaluate their performance by detecting either of them or both.
In clinical routine, MR images are typically acquired using different sequences or weightings in which the tissues appear in specific intensities. Among the most common ones are T1, T2, Fluid-attenuated inversion recovery (FLAIR), or Proton density (PD)-weighting.
In FLAIR images -- a standard protocol for routine clinical imaging in neurology -- lesions are hyperintense compared to the rest of the tissue and also tumors are usually brighter. Because of this, FLAIR images are often used in SAS of brain MRI \cite{BrainLesionAE,AnoVAEGAN,ComparativeStudy,transformer}.

In our work, we leverage this prior knowledge to build a baseline that performs anomaly segmentation of brain MRI via simple thresholding of the input FLAIR image.
In particular, the main contributions of our work are:
\begin{itemize}
  \item We show that learning the distribution of \say{normal} anatomies in FLAIR images using existing autoencoder-based approaches does not provide better segmentation maps of common anomalies in the brain than the input images themselves binarized at a certain threshold intensity.
  \item We provide a simple baseline that requires no learning and outperforms most state-of-the-art SAS methods on common evaluation data sets containing brain tumors and MS lesions.
\end{itemize}

\section{Related Work}
\begin{figure}[htpb]
    \centering
    \includegraphics[width=1.0\textwidth]{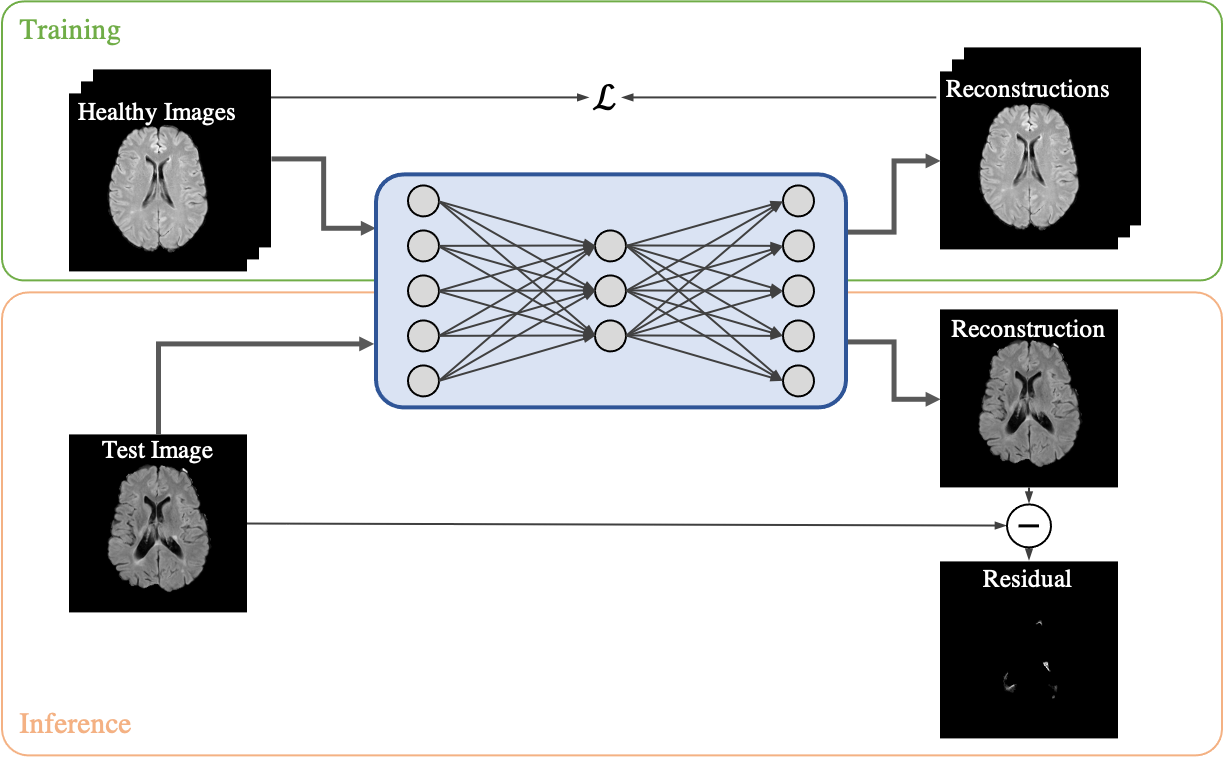}
    \caption{Overview of Autoencoder- and GAN-based SAS. During training, the model learns the distribution of normal anatomies using only images of healthy patients. At inference time, the model generates a "healthy" version of the input image. The anomalies can be determined from the residual image. Image adapted from \cite{ComparativeStudy}.}
    \label{fig:overview_sas}
\end{figure}

Several methods for SAS in brain images have been introduced in recent years. Most of them are based on semi-supervised training of Autoencoders. 
The principle is depicted in Figure \ref{fig:overview_sas}. The model is trained on images without anomalies only to learn a distribution of healthy brain images. During inference, the newly presented image is processed by the model to obtain a \say{healthy} version of the same image. Usually, an anomaly map is then obtained by computing the residual between the input image and its \say{healthy} version. Pixels of the anomaly map above a threshold are then considered anomalous.

In \cite{CTAE}, the authors trained a Bayesian Autoencoder to perform anomaly segmentation on CT images.
Chen and Konukoglu \cite{constrainedAE} built an Adversarial Autoencoder with an additional constraint forcing the input image and its reconstruction to be close in latent space.
Another reconstruction-based technique was proposed in \cite{AnoVAEGAN}, where Baur \textit{et al.} built a VAEGAN to increase reconstruction fidelity and realism of the reconstructed images.
Zimmerer \textit{et al.} \cite{contextVAE,BrainVAE} added gradient information from the loss-function of Variational Autoencoders (VAEs) to the reconstruction error, offering superior anomaly maps.

Restoration methods use the trained model to perform gradient optimization on the input image to construct an image that is both similar to the input and close to the distribution of normal anatomies learned by the model.
Anomaly maps are again computed as the residual between the input- and the optimized image.
An early example of this technique was proposed by Schlegl \textit{et al.} \cite{f-AnoGAN}. They retrieve the closest version to an image that a Generative Adversarial Network (GAN) -- trained on images of healthy patients only -- can produce.
Chen \textit{et al.} \cite{GMVAE} also used restoration by maximizing the evidence lower bound (ELBO) of an image on a Gaussian Mixture VAE (GMVAE).

Recently, Baur \textit{et al.} published a comparative study \cite{ComparativeStudy}, comparing all the methods above on the same data sets with a unified architecture. We use their results in this work to compare our baseline against all of these techniques. We use the same data sets for evaluation and use a similar pre- and post-processing pipeline.
In \cite{bae}, Baur \textit{et al.} proposed to use a U-Net-like Autoencoder with skip-connections and in \cite{ssae}, the same authors introduced a multi-scale Autoencoder utilizing a laplacian pyramid.
While \cite{bae} and \cite{ssae} were both trained on the same data and used identical pre-processing as \cite{ComparativeStudy}, only \cite{ssae} was evaluated on one public data set and can be compared in this work.
Pinaya \textit{et al.} \cite{transformer} achieved impressive results in SAS of brain MRI. They trained a Vector Quantised VAE (VQ-VAE) on a large cohort of FLAIR images of healthy subjects and later trained an ensemble of autoregressive Transformers in its latent space. The Transformers provide an explicit probability distribution of pixels in the latent space. Pixels with low posterior probability are considered anomalous. 
Since this method is not included in the comparative study by Baur \textit{et al.} \cite{ComparativeStudy}, we compare our results to theirs in a separate experiment.

Lastly, anomaly detection was used by van Hespen \textit{et al.} \cite{nature_ad} to detect chronic brain infarcts on MRI. They made a patch-based detection approach using a scoring function based on the latent space distances instead of the reconstructed image.
The anomaly score for the whole image is calculated as a combination of all patches, resulting in a coarse segmentation map.
We did not include this method in our experiments, because the models were trained on non-publicly available data and the model parameters are not open-source.
However, they showed that SAS methods are able to spot unseen anomalies. Their system was able to identify anomalies missed in the annotation of an expert radiologist, proving the usefulness of such approaches.
\section{Experiments}
In the following, we present the data sets we used to evaluate our baseline, pre- and post-processing steps and evaluation metrics.

\subsection{Datasets}
We compare our baseline to all the publicly available data sets used for evaluation in Baur \textit{et al.} \cite{ComparativeStudy} and Pinaya \textit{et al.} \cite{transformer}.

To evaluate brain tumor detection, we use the training set of the 2020 version of the Multimodal Brain Tumor Image Segmentation Benchmark (BraTS) \cite{brats1,brats2,brats3}.
It contains T1, T2, and FLAIR scans of 371 subjects acquired across 19 institutions with multimodal, 3 Tesla MRI scanners. It also contains manual segmentations of the tumor regions by up to four raters. The BraTS images are already skull stripped.
The MSLUB \cite{mslub} data set consists of T1, T2, and FLAIR images of 30 subjects with multiple sclerosis (MS). They have been acquired at the University Medical Center Ljubljana (UMCL) with a 3 Tesla Siemens Magnetom Trio MR system. The consensus of three experts on white matter lesion segmentation is also included.
As in \cite{transformer}, we evaluate on the White Matter Hyperintensities Segmentation Challenge (WMH) \cite{wmh}. For this data set, T1 and FLAIR scans of 60 patients were acquired at three different sites in the Netherlands and Singapore. The sites used 3 Tesla MRI scanners from Philips, Siemens, and GE. Manual segmentation of the lesions was conducted by an expert radiologist.
Lastly, we use the training data of the 2015 Longitudinal MS Lesion Segmentation Challenge \cite{msseg}. This dataset has 21 T1, T2, PD, and FLAIR weighted MRI scans from 5 subjects recorded at the John Hopkins MS Center with a 3 Tesla Philips scanner. Manual lesion segmentations are available from two raters. We use the ratings of rater one (as indicated by the filename "mask1.nii") for our evaluation.

Tumors are usually much larger anomalies than MS lesions. We evaluated the exact distribution of anomaly sizes by performing a 3D connected component analysis on the segmentation maps of all data sets (table \ref{tab:data_stats}). MSLUB has the smallest anomalies and also the largest number of anomalies per scan.

\begin{table}[htpb]
\caption{Results of the 3D connected component analysis of the segmentation maps of all data sets after being registered to SRI space \cite{sri} and binarized with threshold $0.9$ (See section \ref{sec:preprocessing}).}
\label{tab:data_stats}
\centering
\bgroup
\begin{tabular}{lllll}
\toprule
& \textbf{BraTS} & \textbf{MSLUB} & \textbf{WMH} & \textbf{MSSEG2015} \\
\midrule
\textbf{Avg. anomalies per scan} & 5 & 107 & 65 & 35 \\
\textbf{Avg. anomaly size (voxels)} & 18027 & 106 & 194 & 224 \\
\bottomrule
\end{tabular}
\egroup
\end{table}

\subsection{Pre-processing} \label{sec:preprocessing}
\begin{figure}[htpb]
    \centering
    \includegraphics[width=0.7\textwidth]{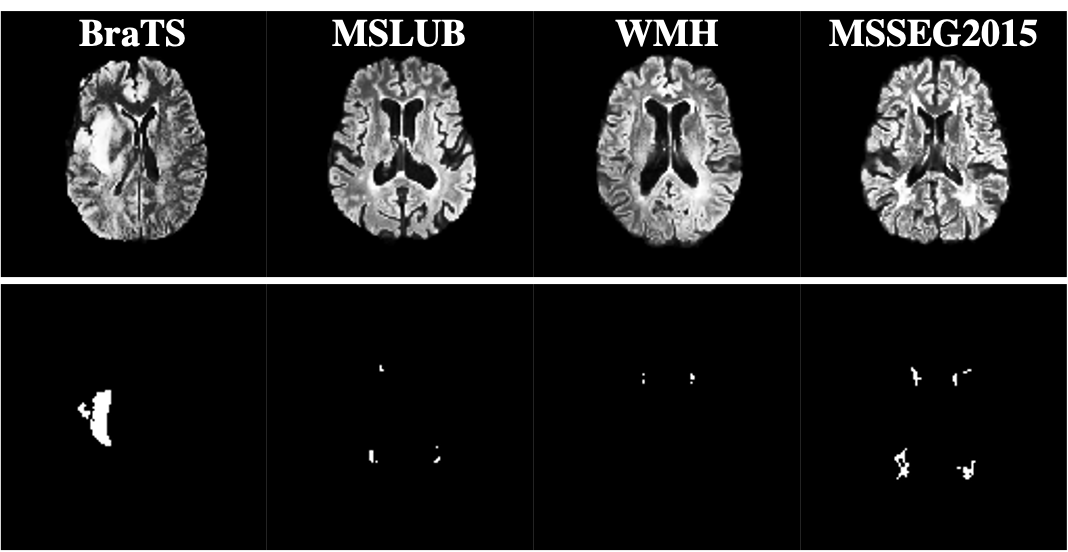}
    \caption{Pre-processed samples and histogram-equalized (top row) and their corresponding ground truth segmentations (bottom row) from the four data sets.}
    \label{fig:dataset_samples}
\end{figure}

Our pre-processing pipeline closely follows Baur \textit{et al.} \cite{ComparativeStudy}. First, we skull strip the FLAIR scans using ROBEX \cite{robex}. Subsequently, we register them to the SRI space \cite{sri}. Specifically, since \cite{sri} does not contain a FLAIR Atlas, we register the T1-weighted images of all data sets and apply the same transformation to the FLAIR images and the ground truth segmentation masks. This is possible, as T1- and FLAIR images and the segmentation files are co-registered in all the data sets used.
Performing registration before skull stripping resulted in failed registrations in early experiments.
The registration step is not vital for our algorithm but was purely done to ensure comparability with other methods.
Figure \ref{fig:dataset_samples} shows samples of pre-processed images from all four data sets.

During the registration process, aliasing effects occur in the -- initially binary -- ground truth segmentation masks that cause these masks to also have non-binary voxel values between $0$ and $1$ after registration. When loading the data, a decision needs to be made at which threshold a voxel in the segmentation map belongs to the segmented region.
We consulted an expert radiologist and visually found $0.4$ to be an acceptable threshold, but finally decided to follow Baur \textit{et al.} \cite{ComparativeStudy} in using $0.9$, to ensure better comparability.
Note that altering this threshold has large effects on the performance of the evaluated models, especially on data sets with many small anomalies like lesions. A low threshold favors models that overestimate the true size of the anomalies, while a high threshold does the opposite.

\subsection{Method}
While other SAS methods usually compute anomaly maps using Neural Networks, we propose to only perform histogram equalization on the pre-processed input images and use the results directly as anomaly maps since lesions and tumors often are hyperintense in FLAIR images anyway. Histogram equalization is necessary to compensate for contrast variations among different scanner types and allows to define a global (or at least dataset-wise) threshold for binarization of the anomaly maps. We used the \verb|equalize_hist| function of scikit-image \cite{skimage} with the default value of 256 bins and a binary mask considering only pixels belonging to the brain and excluding the background.
Using FLAIR images is a fair comparison since Baur \textit{et al.} \cite{ComparativeStudy} and Pinaya \textit{et al.} \cite{transformer} also trained and evaluated on FLAIR images only.
Our method does not require any training data or learning procedure and scales trivially to arbitrary resolutions.

\subsection{Post-processing}
As our only post-processing step, we perform a connected component analysis per scan on the 3D voxels as in \cite{ComparativeStudy} and discard all anomalies with less than 20 voxels. This value was found empirically and causes our algorithm to potentially miss very small anomalies. However, it greatly reduces the noise in the anomaly maps and thereby enhances their readability.

\subsection{Metrics} \label{sec:metrics}
We quantitatively assess the anomaly segmentation performance of our method using a variety of metrics also frequently found in related works. All metrics are produced dataset-wise.
Initially, we compute Precision-Recall curves and report the area under it (AUPRC).
We also provide an upper limit for the Dice similarity coefficient ($\lceil$DSC$\rceil$), computed using a search over $n=100$ thresholds.
Lastly, we also provide the area under the receiver operating characteristics curve (AUROC).
\section{Results} \label{sec:results}

\begin{table}[htpb]
\caption{Comparison of our proposed baseline to selected models of Baur \textit{et al.} \cite{ComparativeStudy} and \cite{ssae}. We used slices 15 to 125 of the registered FLAIR images and a resolution of $128 \times 128$.}
\label{tab:experiment1}
\centering
\bgroup
\begin{tabular}{lllllll}
\toprule
& \multicolumn{3}{c}{\textbf{MSLUB}} & \multicolumn{3}{c}{\textbf{MSSEG2015}} \\
\cmidrule(r{4pt}){1-1} \cmidrule(r{4pt}){2-4} \cmidrule(){5-7}
Method & $\lceil$\textbf{DSC}$\rceil$ & \textbf{AUPRC} & \textbf{AUROC} & $\lceil$\textbf{DSC}$\rceil$ & \textbf{AUPRC} & \textbf{AUROC} \\
\cmidrule(r{4pt}){1-1} \cmidrule(r{4pt}){2-4} \cmidrule(){5-7}
AE (dense) \cite{ComparativeStudy} & 0.271 & 0.163 & 0.794 & 0.185 & 0.080 & 0.879 \\
AE (spatial) \cite{ComparativeStudy,AnoVAEGAN} & 0.154 & 0.065 & 0.732 & 0.106 & 0.037 & 0.781 \\
VAE (rest.) \cite{ComparativeStudy,GMVAE} & 0.333 & \textbf{0.275} & 0.839 & 0.272 & 0.202 & 0.905 \\
GMVAE(rest.) \cite{ComparativeStudy,GMVAE} & 0.332 & 0.271 & 0.836 & 0.280 & 0.199 & 0.909 \\
f-AnoGAN \cite{ComparativeStudy,f-AnoGAN} & 0.283 & 0.221 & 0.856 & 0.342 & 0.255 & 0.923 \\
SSAE(spatial) \cite{ssae} & 0.301 & 0.222 & - & - & - & - \\
Ours & \textbf{0.374} & 0.271 & \textbf{0.991} & \textbf{0.431} & \textbf{0.262} & \textbf{0.996} \\
\bottomrule
\end{tabular}
\egroup
\end{table}

We evaluate our method in two experiments. First, we report the performance when using slices $15$ to $125$ on a resolution of $128 \times 128$ as in the experiments of Baur \textit{et al.} \cite{ComparativeStudy} and \cite{ssae}.
These slices contain most of the brain region in the SRI space \cite{sri} and tests did not show significant differences in the quantitative evaluation compared to the full volumes.
The results of experiment one are shown in table \ref{tab:experiment1}.
Although for \cite{ComparativeStudy} the code is available online, we did not re-train the models but used the values reported in the respective papers because the training data used is not publicly available. We only report the numbers of a subset of the best performing models, the others can be inspected in the original paper.
In our experiments, our proposed baseline outperforms all other methods in terms of DSC and AUROC and is competitive in AUPRC.
While all models in \cite{ComparativeStudy} use a unified architecture, the detailed architecture of \cite{ssae} is unknown, and the two papers report significantly different performances for the same models on the same data sets, indicating volatility of these methods.

\begin{table}[htpb]
\caption{Comparison of our proposed baseline to Pinaya \textit{et al.} \cite{transformer}. We used slices 84, 85, 86, and 87 of the registered FLAIR images and a resolution of $224 \times 224$.}
\label{tab:experiment2}
\centering
\bgroup
\begin{tabular}{llll}
\toprule
\textbf{} & \multicolumn{3}{c}{\textbf{$\lceil$DSC$\rceil$}} \\
\cmidrule(r{4pt}){1-1} \cmidrule(){2-4}
\textbf{Method} & \textbf{BraTS} & \textbf{MSLUB} & \textbf{WMH} \\
\cmidrule(r{4pt}){1-1} \cmidrule(){2-4}
Transformer \cite{transformer}  & \textbf{0.759}  & 0.465 & 0.441 \\
Ours & 0.738 & \textbf{0.613} & \textbf{0.557} \\
\bottomrule
\end{tabular}
\egroup
\end{table}

In our second experiment, we compare to Pinaya \textit{et al.} \cite{transformer} at a resolution of $224 \times 224$.
In this experiment, there are some differences regarding pre- and post-processing. Pinaya \textit{et al.} \cite{transformer} evaluate on data that was not skull stripped, except for BraTS. They also did not perform any post-processing on the BraTS data set.
They registered to MNI space that has 189 slices instead of SRI with 155 slices. We therefore used slices 84, 85, 86, and 87 instead of 89 to 92 to still ensure a fair comparison.
Lastly, they used the older 2017-version of the BraTS dataset, whereas we used the latest 2020-version.
Our baseline outperforms the Transformer strongly on the MSLUB and WMH data sets and performs only slightly worse on the BraTS data set.
%The result of all metrics of our baseline on all data sets can be found in the supplementary material.

\begin{figure}[htpb]
    \centering
    \includegraphics[width=1.0\textwidth]{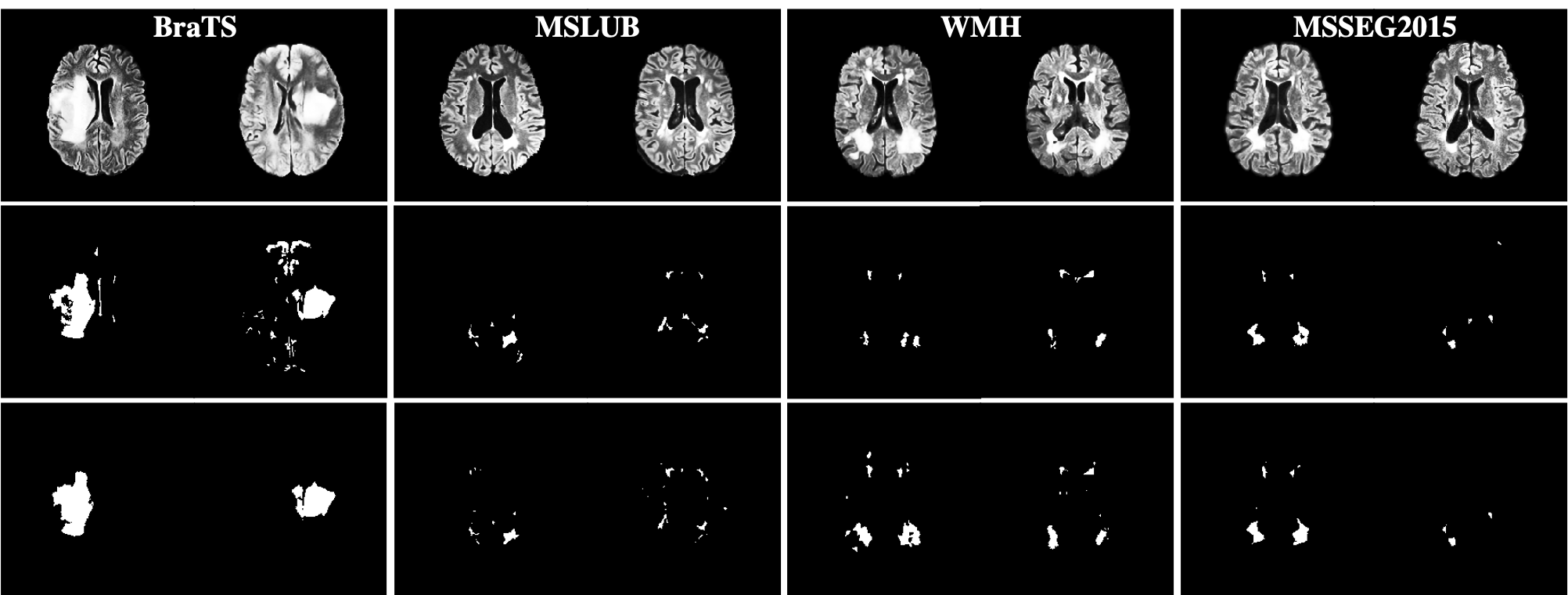}
    \caption{Qualitative results of our baseline. Two samples are shown for each data set. Top row: input image. Middle row: predicted anomaly map, binarized using the threshold that yields the best DSC for each data set. Bottom row: ground truth anomaly segmentation.}
    \label{fig:my_results}
\end{figure}

Figure \ref{fig:my_results} shows the qualitative results of our proposed baseline. The visual segmentation quality based on image-hyperintensities is decent and shows the approximate localization of anomalies.

We also present the quantitative results of the two experiment settings for all data sets using all metrics in table \ref{tab:all_results}. In experiment one, our proposed method performs best on the BraTS data set which has the largest anomalies, and worst on MSLUB with the smallest anomalies. This can partly be attributed to our post-processing where we discard connected components with less than 20 voxels. Datasets with smaller anomalies are more affected by this.
Also in experiment two, BraTS is the data set with the highest $\lceil$DSC$\rceil$ and AUPRC.

\begin{table}[htpb]
\caption{Full results of our proposed baseline on the two experimental settings. Experiment I: using slices 15 to 125 of the registered FLAIR images and a resolution of $128 \times 128$. Experiment II: using slices 84, 85, 86, and 87 of the registered FLAIR images and a resolution of $224 \times 224$.}
\label{tab:all_results}
\centering
\bgroup
\begin{tabular}{llll}
\toprule
& \textbf{$\lceil$DSC$\rceil$} & \textbf{AUPRC} & \textbf{AUROC} \\
\midrule
\multicolumn{4}{l}{Experiment I} \\
\midrule
\textbf{BraTS} & 0.666 & 0.671 & 0.988 \\
\textbf{MSLUB} & 0.374 & 0.278 & 0.991 \\
\textbf{WMH} & 0.457 & 0.339 & 0.979 \\
\textbf{MSSEG2015} & 0.431 & 0.262 & 0.996 \\
\midrule
\multicolumn{4}{l}{Experiment II} \\
\midrule
\textbf{BraTS} & 0.738 & 0.762 & 0.985 \\
\textbf{MSLUB} & 0.613 & 0.571 & 0.993 \\
\textbf{WMH} & 0.557 & 0.504 & 0.984 \\
\textbf{MSSEG2015} & 0.593 & 0.536 & 0.996 \\
\bottomrule
\end{tabular}
\egroup
\end{table}
\section{Discussion}
\begin{figure}[htpb]
    \centering
    \includegraphics[width=1.0\textwidth]{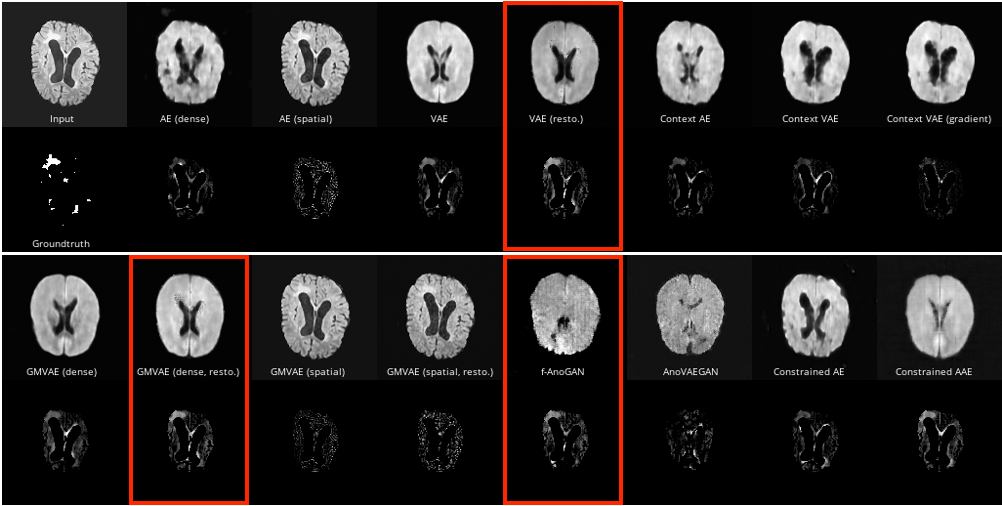}
    \caption{Reconstructions (top row) and residuals (bottom row) of different ML-based SAS techniques. The best performing models are highlighted in red. Image from and best viewed in \cite{ComparativeStudy}.}
    \label{fig:comparative_study_mslub}
\end{figure}

The results in Section \ref{sec:results} show that a simple baseline can outperform or compete with even the strongest related Machine Learning (ML) techniques.
These findings challenge the effectiveness of current ML approaches for SAS.
The results of Baur \textit{et al.} \cite{ComparativeStudy} also show that DSC does not correlate well with reconstruction quality. Especially, one can see in Figure \ref{fig:comparative_study_mslub}, that the best performing models (VAE with restoration, dense GMVAE with restoration, and f-AnoGAN) produce very textureless reconstructions. They can detect the largest connected anomaly located at the dorsal aspect of the right lateral ventricle (note that the images are oriented such that the patients' right ventricle is on the left side of the image) only because it is hyperintense in the input image. We refer to the original paper for a higher-resolution version of this figure.
Hence, we hypothesize that the models in Baur \textit{et al.} \cite{ComparativeStudy} do not perform anomaly segmentation by learning the normal anatomy of the data, but that the necessary information to perform anomaly segmentation with the performance presented in our work is already present in the input image.
The quantitative evaluation of our experiments indicates that using the residual between the model output and the input image actually degrades the segmentation quality of the resulting anomaly map.

While we are aware that our baseline can only detect anomalies that are hyperintense, we argue that other techniques -- especially those using residual maps between the input image and a reconstructed or restored version as anomaly maps -- are not assumption-free, but also impose strong biases on the types of anomalies they can detect. For example, Alzheimer's disease, where one of the symptoms is atrophy of regions of the brain, cannot reliably be detected using pixel-wise residuals.
Some of the existing works \cite{transformer,ComparativeStudy} leverage the same prior knowledge by considering only positive residuals as anomalies for MS lesions. However, our approach appears to make better use of this knowledge.

We point out that there exist anomaly segmentation methods like \cite{nature_ad} that have shown to be able to detect anomalies that are not necessarily hyperintense. These methods, however, do not base their anomaly score on the reconstruction error but have other inductive biases. Van Hespen \textit{et al.} \cite{nature_ad} limit the receptive field of their model with the patch size used.

\section{Conclusion}
In this work, we advanced the current state-of-the-art in SAS of brain MRI by introducing a simple method that requires no learning. Our findings challenge the effectiveness of existing ML-based SAS approaches.
While our work outperforms competing methods, the results still lack behind the ones of expert radiologists and supervised methods presented in \cite{wmh}, \cite{msseg} and \cite{brats3}. This provides evidence for the need to explore alternative methods that overcome current limitations.
These could include new scoring functions or multi-modal approaches. We also encourage the use of prior knowledge to build these models. While this seems counter-intuitive at first -- given the promise of SAS being able to detect any kind of anomalies -- we argue that current methods are also severely limited by their scoring functions in the types of anomalies they are theoretically able to detect.
To this regard, we will explore the use of artificial anomalies in anomaly segmentation.
We hypothesize that through careful creation and selection of artificial anomalies, models can generalize to real anomalies.
Our work also highlights the requirement for a benchmark data set to better compare different techniques against each other.
This benchmark should contain relevant real-world anomalies of brain MRI, but should also not be sufficiently solved via non-ML methods.
Another disadvantage of the presented models is their limited spatial scope. Current SAS methods process the 2D slices of a 3D volume individually. We suspect that making better use of the 3D information of MRI will improve the anomaly detection performance of the models.
We plan to explore the use of 3D machine learning models in future work as they can fully incorporate 3D information, while humans can only process volumes -- such as MRI -- slice-wise.
%
%
% ---- Bibliography ----
%
% BibTeX users should specify bibliography style 'splncs04'.
% References will then be sorted and formatted in the correct style.
%
\bibliographystyle{splncs04}
\bibliography{references}
\end{document}